\documentclass[iop]{emulateapj}
\newcommand{\possessivecite}[1]{\citeauthor{#1}'s \citeyear{#1}}

\usepackage{pdfpages}
\usepackage{amsmath}

\shorttitle{The history of star formation in the 30~Doradus region: Hodge~301} \shortauthors{Cignoni et al.}

\begin{document}


\title{Hubble Tarantula Treasury Project V. The star cluster Hodge
  301:\\ the old face of 30~Doradus}%

\author{ M. Cignoni\altaffilmark{2,3,4}, E. Sabbi\altaffilmark{3},
  R. P. van der Marel\altaffilmark{3}, D. J. Lennon\altaffilmark{5},
  M. Tosi\altaffilmark{6},E. K. Grebel\altaffilmark{7}, Gallagher,
  J. S., III\altaffilmark{8}, A. Aloisi\altaffilmark{3},G. de
  Marchi\altaffilmark{9},D. A. Gouliermis\altaffilmark{10,11},
  S. Larsen\altaffilmark{12}, N. Panagia\altaffilmark{3,13,14},
  L. J. Smith\altaffilmark{15}} \email{michele.cignoni@unipi.it}

\altaffiltext{1}{Based on observations with the NASA/ESA Hubble Space Telescope, obtained at the Space Telescope Science Institute, which is operated by AURA Inc., under NASA contract NAS 5-26555}
\altaffiltext{2}{Department of Physics - University of Pisa, Largo Pontecorvo, 3 Pisa, 56127, Italy }
\altaffiltext{3}{Space Telescope Science Institute, 3700 San Martin Drive, Baltimore, MD, 21218, USA }
\altaffiltext{4}{INFN, Largo B. Pontecorvo 3, 56127 Pisa, Italy}
\altaffiltext{5}{European Space Astronomy Centre, Apdo. de Correo 78,
  E-28691 Villanueva de la Canada, Madrid, Spain}
\altaffiltext{6}{INAF-Osservatorio Astronomico di Bologna, Via Ranzani 1, I-40127
Bologna, Italy}
\altaffiltext{7}{Astronomisches Rechen-Institut, Zentrum f\"ur Astronomie der
Universit\"at Heidelberg, M\"onchhofstr. 12-14, 69120 Heidelberg, Germany}
\altaffiltext{8}{Department of Astronomy, University of Wisconsin-Madison, WI 53706,
USA}
\altaffiltext{9}{European Space Research and Technology Centre,
  Keplerlaan 1, NL-2200 AG Noordwijk, the Netherlands}
\altaffiltext{10}{Zentrum f\"ur Astronomie der Universit\"at Heidelberg, Institut f\"ur Theoretische Astrophysik, Albert-Ueberle-Str.\,2, 69120 Heidelberg, Germany }
\altaffiltext{11}{Max Planck Institute for Astronomy, Konigstuhl 17,
  D-69117 Heidelberg, Germany}
\altaffiltext{12}{Department of Astrophysics, Radboud University, PO Box 9010, NL-6500 GL Nijmegen, the Netherlands}
\altaffiltext{13}{INAF-NA, Osservatorio Astronomico di Capodimonte,
  Salita Moiariello 16, I-80131 Naples, Italy}
\altaffiltext{14}{Supernova Ltd, OYV 131, Northsound Rd., Virgin Gorda
  VG1150, Virgin Islands, UK} 
\altaffiltext{15}{European Space Agency and Space Telescope Science Institute, 3700 San
Martin Drive, Baltimore, MD 21218, USA}

\begin{abstract}

  Based on color-magnitude diagrams (CMDs) from the Hubble Space
  Telescope Hubble Tarantula Treasury Project (HTTP) survey, we
  present the star formation history (SFH) of Hodge~301, the oldest
  star cluster in the Tarantula Nebula. The HTTP photometry extends
  faint enough to reach, for the first time, the cluster pre-main
  sequence (PMS) turn-on, where the PMS joins the main sequence. Using
  the location of this feature, along with synthetic CMDs generated
  with the latest PARSEC models, we find that Hodge~301 is older than
  previously thought, with an age between 26.5 and 31.5 Myr. From this
  age, we also estimate that between 38 and 61 supernovae Type-II
  exploded in the region. The same age is derived from the main
  sequence turn-off, whereas the age derived from the post-main
  sequence stars is younger and between 20 and 25 Myr. Other relevant
  parameters are a total stellar mass of
  $\approx 8800\,\pm 800$M$_{\odot}$ and average reddening E(B$-$V)
  $\approx 0.22-0.24$ mag, with a differential reddening
  $\delta$E(B$-$V)$\approx 0.04$ mag.

\end{abstract}





\section{Introduction}
 
The closest starburst region, 30~Doradus in the Large Magellanic Cloud
(LMC), allows one to study the star formation (SF) process on a
variety of scales, from the dense cluster R~136, possibly a newborn
globular cluster, via the massive star-forming region NGC~2070, the
``old'' (20 to 25 Myr, \citealt{grebel00}) populous cluster Hodge~301,
to the many diffuse star-forming regions like NGC~2060. Deciphering
the history of 30~Doradus is therefore a unique opportunity to
understand how star formation originates and propagates. In previous
work \citep{cigno15} we studied NGC~2070, a giant region already
active 7 Myr ago and possibly up to 20 Myr ago. Here we analyse Hodge
301 ($\alpha_{2000} = 05^h38^m16^s, \delta_{2000} = -69\degr03'58''$),
one of the oldest structures in 30~Doradus, approximately located at 3
arc-minutes ($\sim 44$ pc, assuming a distance of 50 kpc) to the
northwest of R~136 (\citealt{ho88}).

Hodge~301 (hereinafter H301) was previously studied by
\cite{mendoza73}, \cite{mcgregor81}, \cite{melnick85},
\cite{lortet91}, \cite{walborn97} (hereinafter WB97) and
\cite{grebel00} (hereinafter GC00), among others. As a part of
an extensive optical spectral classification effort of the stellar
populations within the 30~Doradus Nebula, WB97 classified H301 as a
cluster in the \emph{h} and $\chi$ Persei phase (namely containing A-
and M-type supergiants), with an age $\sim 10$ Myr, the oldest in the
30~Doradus complex. GC00 used a synergy between spectroscopy,
Ultraviolet Imaging Telescope (UIT) photometry and deep WFPC2/HST
photometry reaching V$\approx 24$ to study age, initial mass function
(IMF), and reddening of H301. By comparing the loci of main-sequence
turn-off (MSTO) and post-MS stars with Padova and Geneva stellar
models, they derived an age of 20-25 Myr. Concerning the IMF, for the
mass range 1.26$-$10$\,$M$_{\odot}$ they derived a slope close to
Salpeter. Finally, using the UIT photometry they found a mean
reddening E(B$-$V)=$0.28\pm 0.05$.

More recently, using spectroscopy and a qualitative comparison with
the evolutionary models of \cite{brott11}, \cite{evans15} found an age
of 15$\pm$5 Myr.

\begin{figure*}[t]
\centering \includegraphics[width=13cm]{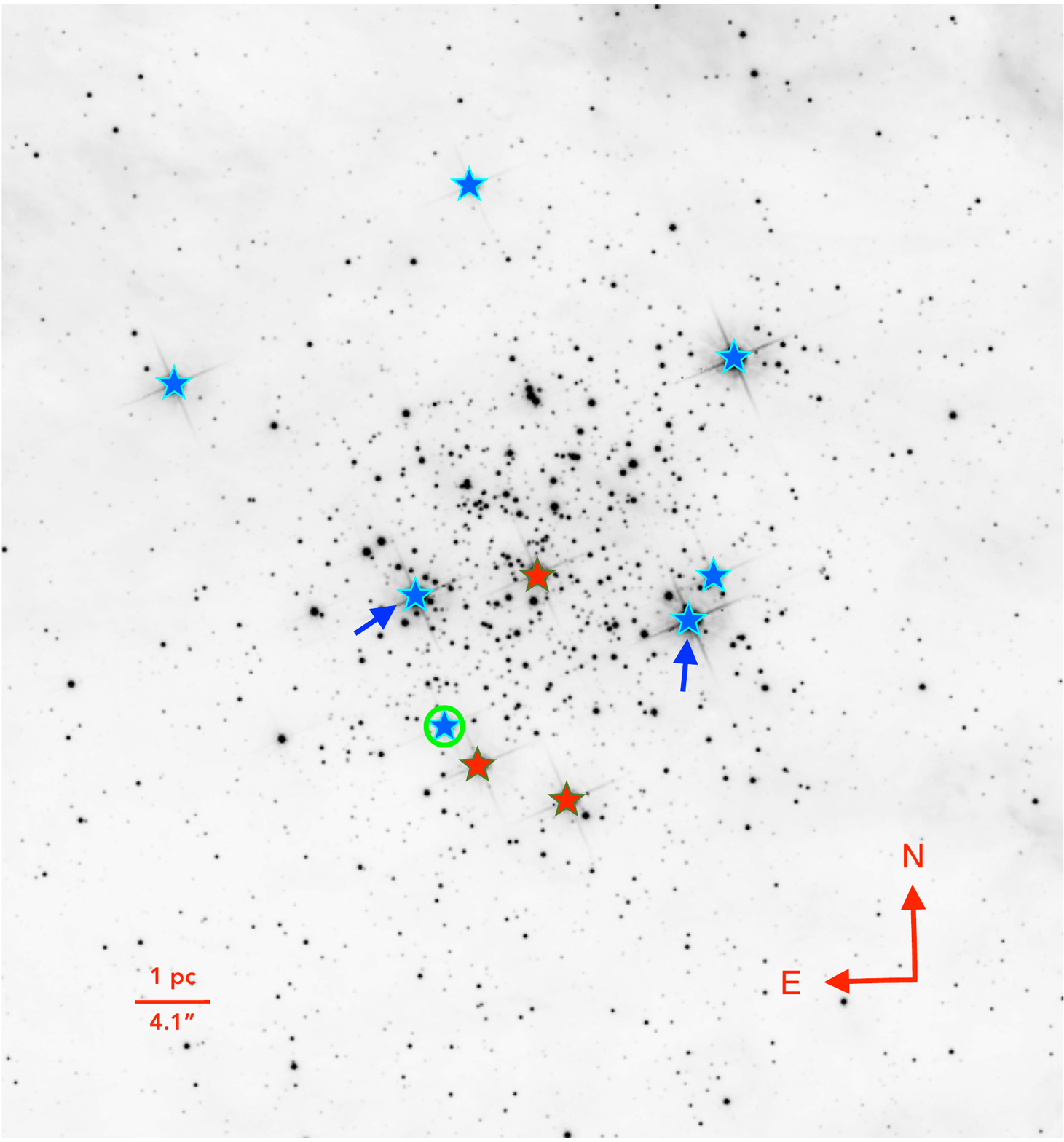}
\caption{Inverted grey-scale image (F555W) of Hodge~301. Blue
  (F555W-F775W$<1$) and red (F555W-F775W$>1$) supergiant stars
  (F555W$<15$) are indicated with blue and red filled starred symbols,
  respectively. The blue arrows point to the two BSGs discussed in the
  present analysis (see text). The open green circle indicates a
  spectroscopic binary (\citealt{walborn97}) that may have a compact
  companion. }
\label{chart} 
\end{figure*}

In this paper we rederive H301's age using the photometric
capabilities of the survey HTTP\footnote{The HTTP Photometric Catalog
  can be downloaded at
  https://archive.stsci.edu/prepds/30dor/Preview/observations.html. }
(\citealt{sabbi12, sabbi15}). Figure \ref{chart} shows a F555W-band
inverted grey-scale image of the cluster. The depth of our CMDs
(V$\approx 26$) allows us for the first time in this cluster to reach
the magnitude of the PMS turn-on (TOn; V$\approx 24-25$), the point of
the color-magnitude diagram (CMD) where the PMS stars join the
MS. This stellar feature is used to measure the cluster age in a way
that is independent of the previous analysis, mostly based on the MSTO
and post-MS stars. For this task we used the synthetic CMD approach
which allows us to fit several crucial features of the CMD (MSTO, PMS
TOn and field contamination) simultaneously. The advantage with
respect to the classical isochrone fitting is the full consideration
of evolutionary times, magnitude and color spreads from photometric
errors, incompleteness, and unresolved binaries.

The structure of the present paper is as follows. In Section 2 we
present the data and we discuss the relevant stellar phases of H301's
CMD. In Section 3 we construct a library of synthetic CMDs based on
stellar isochrones and we use them to locate the MSTO and TOn in the
data. In Section 4 we recover the most likely history for
H301. Conclusions close the paper.


\section{Data}

The survey HTTP has gathered unprecedented photometric data with the
Hubble Space Telescope (HST) over the entire Tarantula Nebula in the
near UV (WFC3/UVIS F275W and F336W), optical (ACS/WFC F555W and
F658N), and near IR (WFC3/IR F110W and F160W). Here we focus on the
optical CMD of H301, the only one deep enough to reach 30 Myr old PMS
stars.

\begin{figure}[t]
\centering \includegraphics[width=9cm]{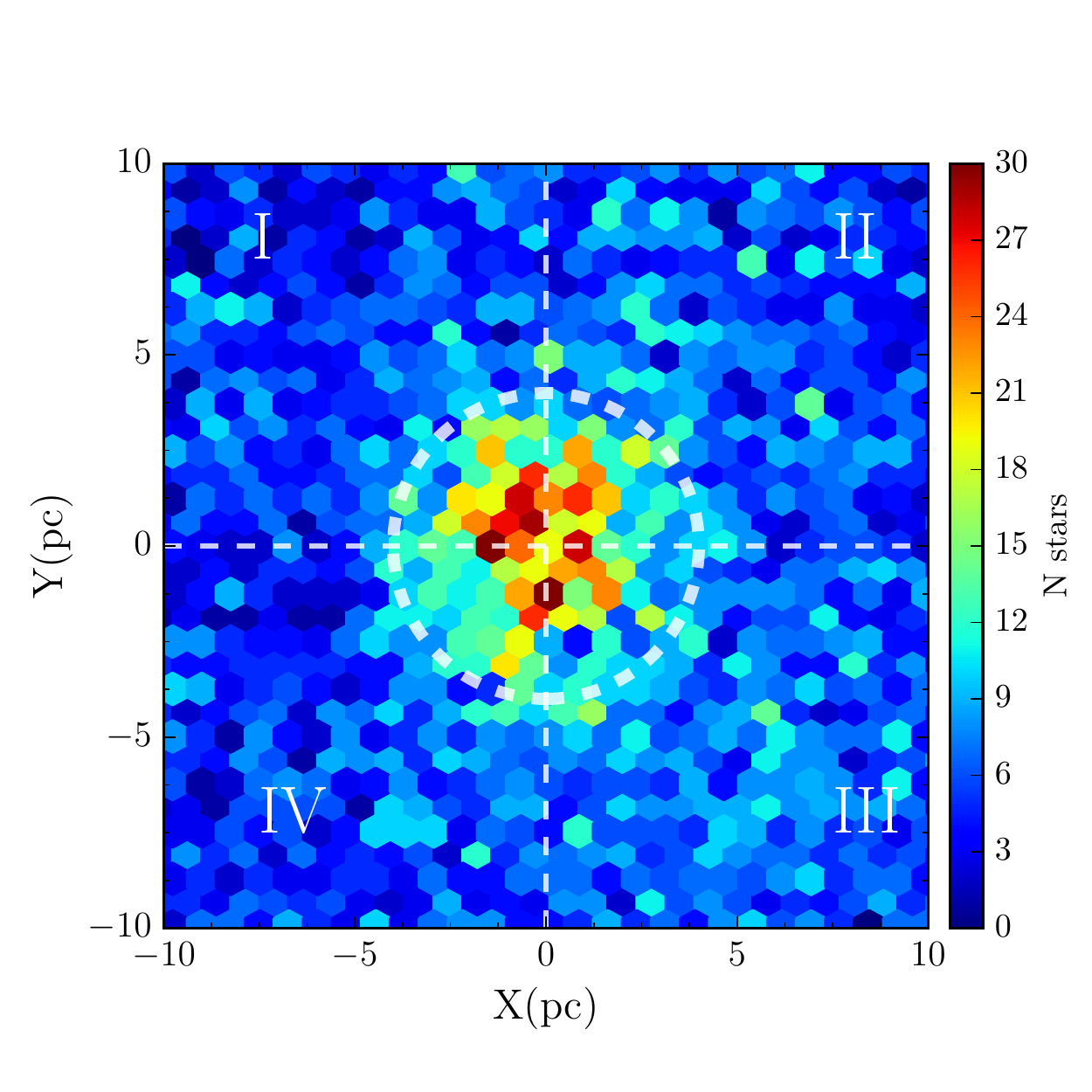}
\caption{Spatial distribution of H301. 1 pc corresponds to $\approx 4$
  arcseconds. The central hole is an artifact of the incompleteness,
  due to the severe central crowding. The dashed circle indicates a 4
  pc radius.}
\label{map} 
\end{figure}

Figure \ref{map} shows the density map of the 10 pc region around
H301. The cluster centroid was chosen as the point minimizing the
distance to all stars.

In order to determine the cluster members, we found the radius at
which the stellar density drops to a value indistinguishable from the
field. Stars within this radius are assumed to belong to the cluster,
and those outside are treated as belonging to the field population.

Figure \ref{profiles} shows the radial profile (number of stars per
pc$^2$) of H301 as calculated in the quadrants of Fig. \ref{map} (the
top-left, top-right, bottom-right, and bottom-left profiles are
indicated with green, blue, red, and black slopes,
respectively). Despite the slight asymmetry (the top-left profile
shows an excess of stars between 1 and 2 pc from the cluster center),
we found that a 4 pc radius encloses about 85\% of the stars of H301,
minimizing at the same time field contamination. Figure \ref{cmd}
shows the corresponding CMD. Photometric errors, as derived from
artificial stars tests (see further down in this Section), are also
shown on the right side. Important features of this CMD are:

\begin{figure}[t]
\centering \includegraphics[width=9cm]{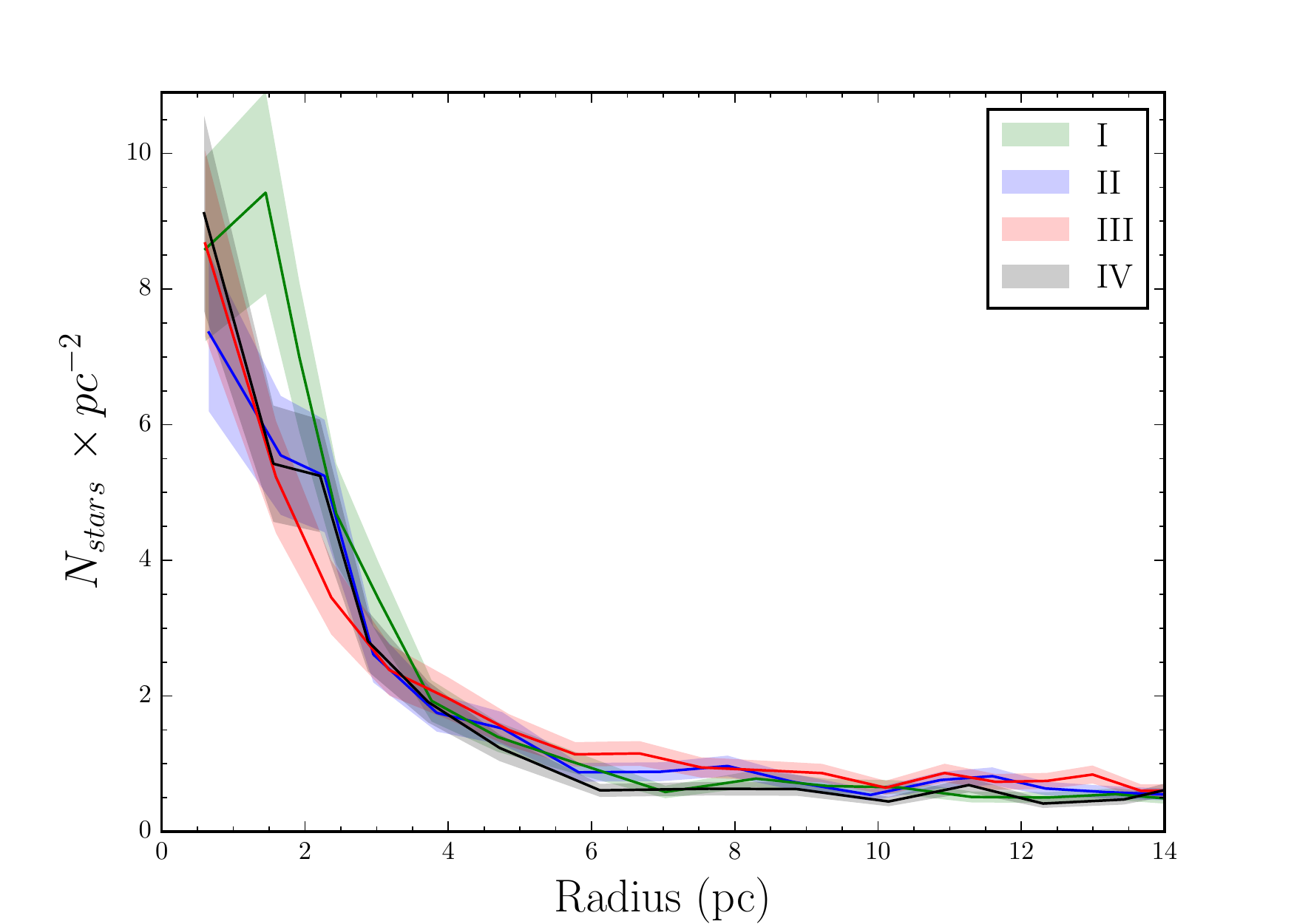}
\caption{Radial profiles of H301 in four quadrants (see
  Fig. \ref{map}). The shadowed areas show the Poissonian
  uncertainties. The differences in one quadrant (green) may indicate
  that the cluster is not spherical.}
\label{profiles} 
\end{figure}

\begin{figure}[t]
\centering \includegraphics[width=8cm]{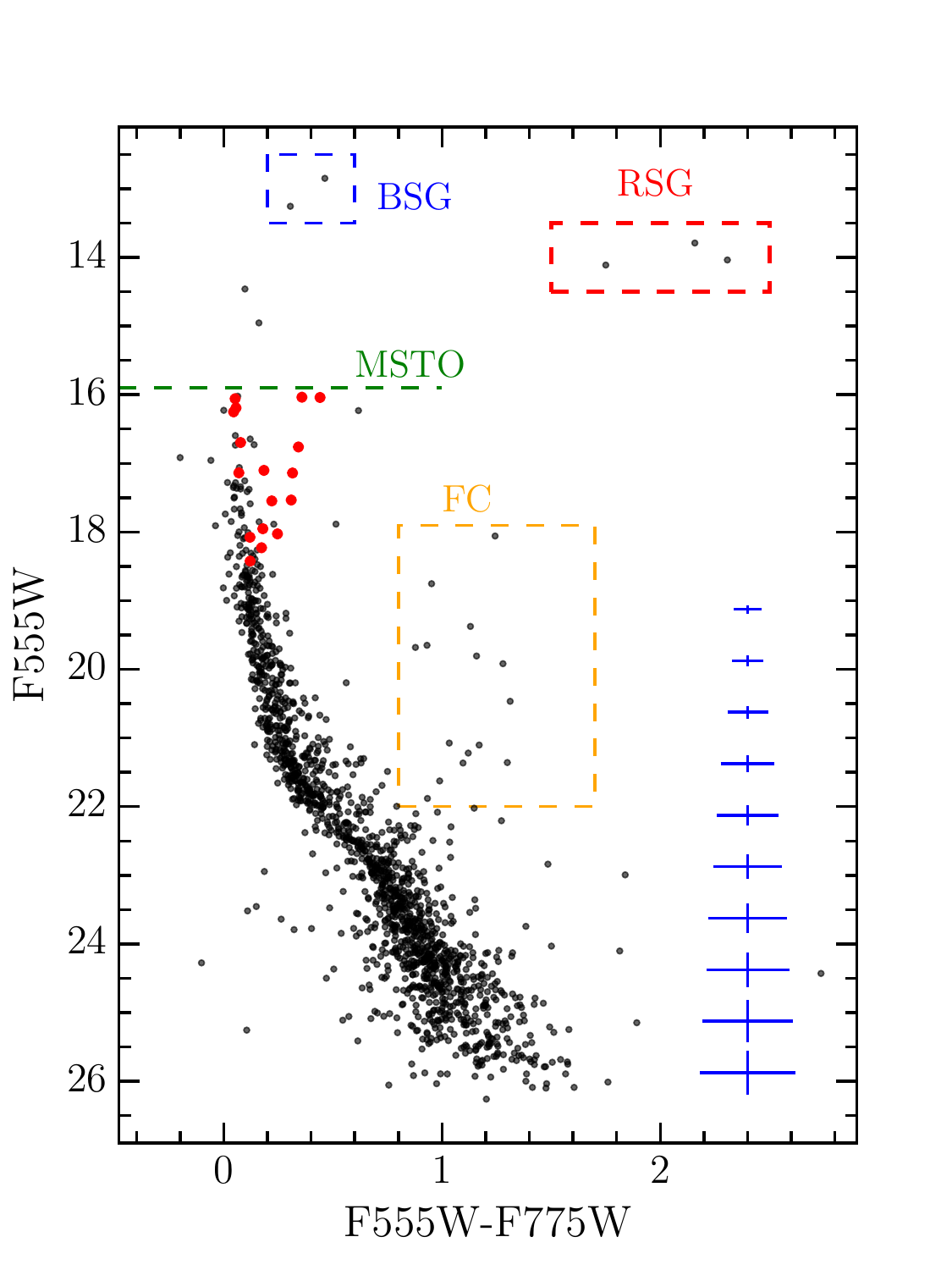}
\caption{F555W vs F555W-F775W CMD of H301 within 4 pc from the cluster
  center. Relevant stellar species and contamination are
  highlighted. Red circles indicate stars brighter than F555W$\sim 19$
  characterized by H$\alpha$ excess.}
\label{cmd} 
\end{figure}

\begin{figure}[t]
\centering \includegraphics[width=9cm]{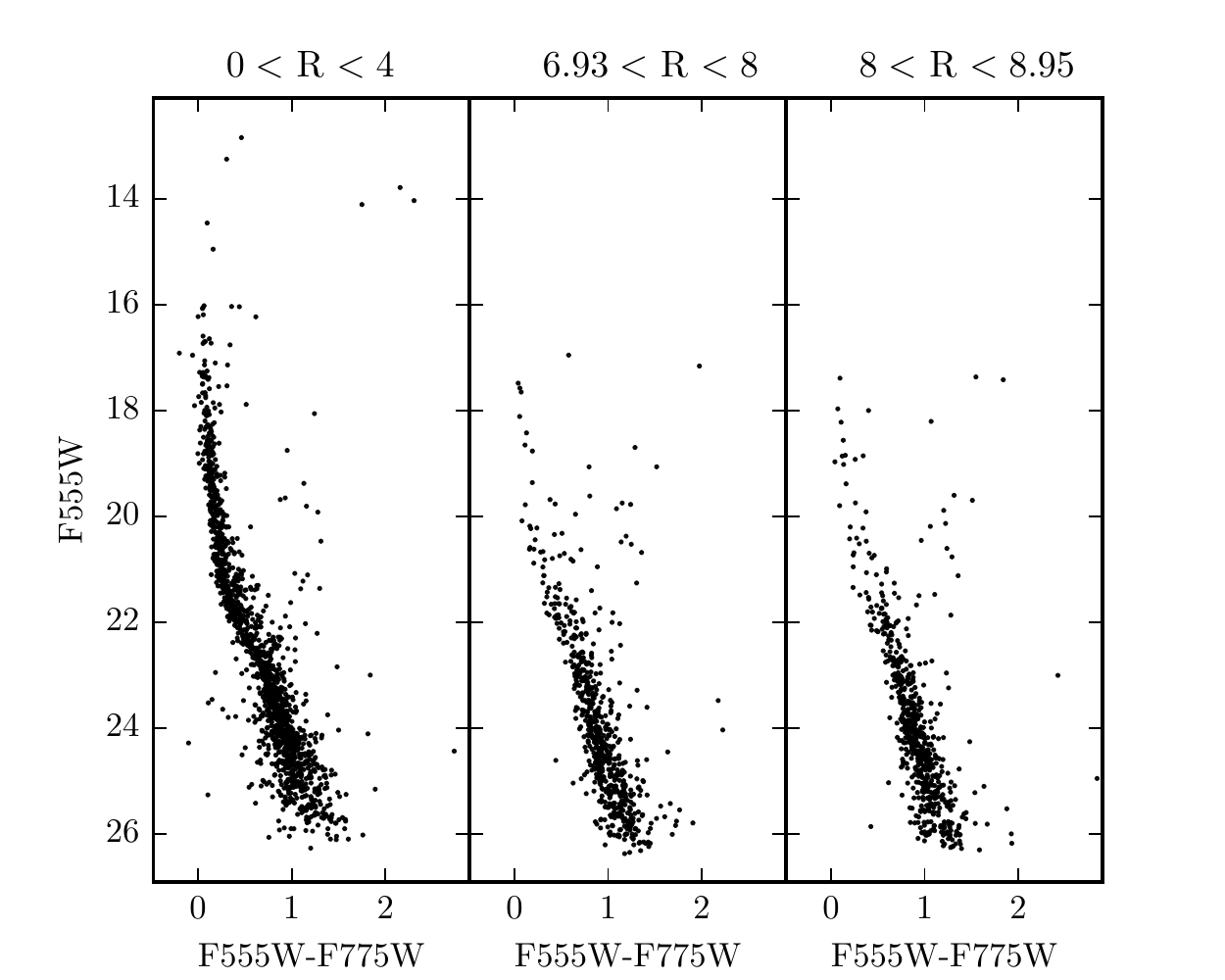}
\caption{From left to right, CMD of stars between 0 and 4 pc, 6.93 and
  8 pc, 8 and 8.95 pc from the center of H301.}
\label{field} 
\end{figure}

\begin{itemize}
\label{bes}
\item An extended MS ranging from F555W$\approx$16 to
  F555W$\approx$26.  For stars in the magnitude range $16<$F555W$<18$,
  the MS is broader than expected on the basis of photometric errors
  (see figure). As already discovered by GC00 (see their Fig. 6), the
  MS is likely broadened toward the red by the presence of Be
  stars. Concerning these objects, GC00 also showed that they are most
  common among the early B-type stars and show the largest Balmer and
  IR excess for the early types (see also \citealt{grebel1997}). In
  our photometry, 17 stars above F555W$\sim 19$ show an H$\alpha$
  excess\footnote{Measured with respect to the median color
    F555W-F658W in the F555W-F658W vs F555W-F775W diagram (see
    \citealt{demarchi10} for details on this approach).} (filled red
  circles in the figure), as expected in Be-stars. Despite this
  spread, the clear drop of star counts brighter than F555W$\sim$16
  suggests that the MSTO (green dashed line) is not brighter than this
  magnitude;

\item A group of stars up to three magnitudes brighter than the MSTO;
  the two brightest and the three reddest are presumably He-burning
  stars of intermediate mass on the blue (blue super giant, BSG) and
  red (red super giant, RSG) side of the blue loop (BL),
  respectively. Their membership is certain, as indicated by the plots
  in Fig. \ref{field}, where the CMDs of stars in three different
  annuli (panels from left to right show stars between 0 and 4 pc,
  6.93 and 8 pc, 8 and 8.95 pc from the cluster center) of equal area
  around the center of H301 are compared. The two outer annuli
  represent pure field samples and only a few upper-MS stars populate
  their CMDs above F555W$\sim 20$. In addition, no upper-MS star above
  F555W$\sim 17$ is detected. On the other hand below F555W$\sim 22$
  field contamination increases significantly.

\item A group of stars at the right of the MS (indicated with ``FC'',
  field contamination, in Figure \ref{cmd}). As suggested by the CMDs
  of the two outer fields in Figure \ref{field}, the entirety of these
  stars is compatible with being red giant (RGB) and red clump (RC)
  stars from the field of the LMC. The elongated shape of the field RC
  (visible in the CMD as the over-density around F555W$\approx 20$ and
  color around 1.2$-$1.4) suggests the presence of some differential
  reddening.

\end{itemize}

The presence of two blue stars above the MSTO around V$\sim$14-15
magnitudes reflects the well-known problem that while theoretical CMDs
predict a post main sequence gap (PMSG) between the end of the main
sequence and the presumed core He burning A-supergiants (the two stars
near V$\sim$13 magnitudes), this gap is not observed.  In clusters of
this approximate age the PMSG is populated by bright B-type giants and
supergiants (of similar color to the main sequence dwarfs and
sub-giants), in the case of Hodge 301 these two stars are confirmed
cluster members with spectral types B3\,Ib and B2\,II-III(n)e (see
\citealt{evans15}). While the fact that the precursor of SN1987A was a
B3\,I star (\citealt{walborn89}) indicates that core He burning stars
can inhabit this part of the CMD/Hertzsprung Russel diagram (HRD)
there is as yet no unambiguous method for distinguishing between core
H and He burning blue stars (see \citealt{hunter08}, \citealt{vink10},
\citealt{grebel1996}, and \citealt{mcevoy15} for discussions of how
rotational velocity distributions, mass-loss considerations and binary
frequency might define the width of the main sequence for B-stars in
the LMC). For now we will assume that the MSTO is as shown in
Fig. \ref{cmd} but will return to this point in the discussion.

\begin{figure}[t]
\centering \includegraphics[width=9cm]{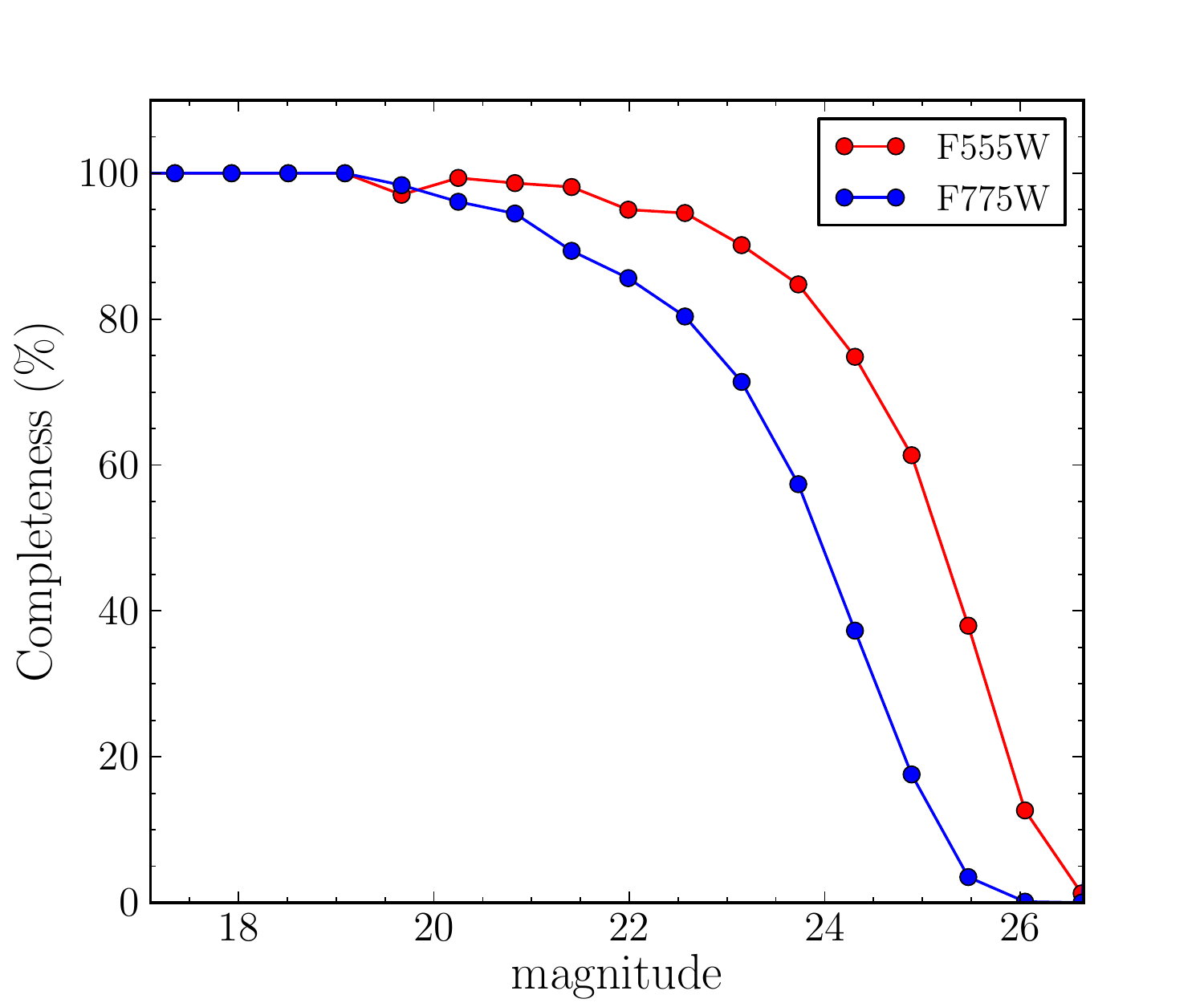}
\caption{Average photometric completeness in F555W (red symbols) and
  F775W (blue symbols).}
\label{perc} 
\end{figure}

In the next Sections we study the CMD of H301 using the synthetic CMD
technique (see, e.g., \citealt{cigno15}).  A mandatory ingredient of
this approach is to accurately test photometric errors and
incompleteness of the data. Here these tests are conducted following a
two step procedure: 1) ``fake'' sources are injected (one at a time)
following a uniform distribution onto the actual images. The source
detection routine used for our science images is applied to the fields
containing the combined actual images and the fake sources. Counting
how many fake stars are lost as a function of magnitude and position
provides the map of the local incompleteness. Note that if the latter
is averaged over the entire cluster, the result does not represent the
``true'' average incompleteness, because the distribution of real
stars is not uniform; 2) the local incompleteness is used to restore
the real profile of the cluster (before the incompleteness). Fake
stars are now injected (one at a time) onto the actual images
following this profile and the source detection routine is applied
again. Although the resulting incompleteness is locally identical to
the incompleteness of step 1, its average over the entire cluster is
an unbiased estimate of the ``true'' average incompleteness. Figure
\ref{perc} shows the average completeness level for the filters F555W
and F775W.

\section{Synthetic CMDs}
\label{synth}
Synthetic CMDs are generated using the latest PARSEC
(\citealt{bressan12,tang14}) isochrones, and assuming a Kroupa IMF and
metallicity Z=0.008 (from the mean [Fe/H] for LMC Cepheids of
\citealt{luck98}, referred to the updated \citealt{caffau11}
mixture). 30\% of synthetic stars are considered members of binary
systems and their flux is combined with a secondary star sampled from
the same IMF. To mimic the observational process, each synthetic CMD
is then convolved with photometric errors (derived from the cumulative
distribution of mag$_{\mathrm{out}}$-mag$_{\mathrm{input}}$ of fake
stars) and incompleteness as derived in the previous Section.


We have compared our synthetic CMDs with the massive stars and
low-mass ones of H301.

\begin{figure}[t]
\centering \includegraphics[width=9.5cm]{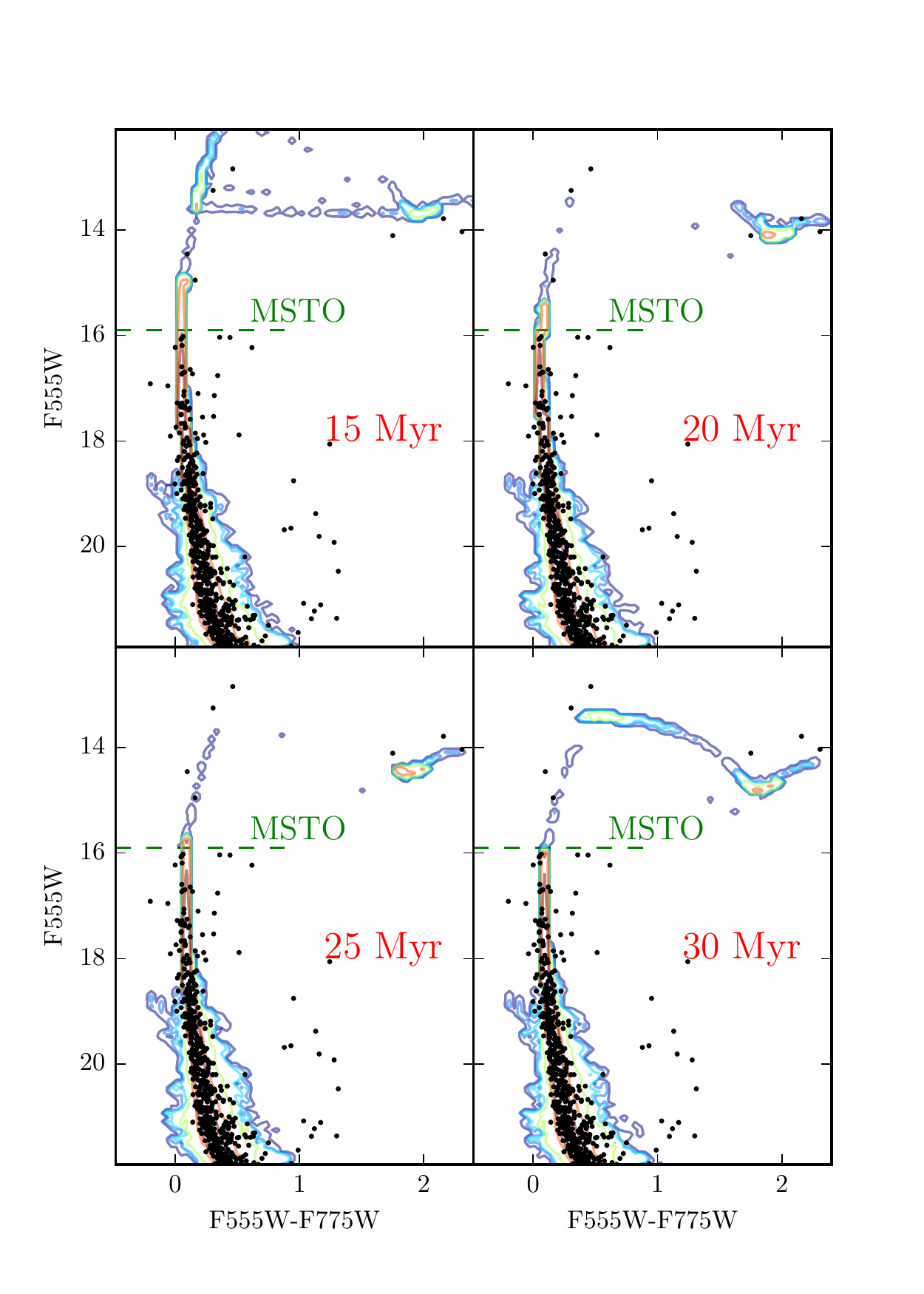}
\caption{Synthetic CMDs (colored contours) for populations of the
  labeled ages, taking unresolved binaries and photometric errors into
  account, overlaid on the F555W vs F555W$-$F775W CMD (dots) of
  H301. }
\label{4cmd} 
\end{figure}

\emph{Massive stars:} Given the likely range of ages of H301, 10-30
Myr, MSTO and BL phases are populated by stars more massive than
$7\,$M$_{\odot}$. Figure \ref{4cmd} shows four synthetic populations
of the labeled ages (15, 20, 25 and 30 Myr) and a duration of star
formation of 1 Myr overlaid to the observations. In order to increase
the visibility of the models in the fastest evolutionary phases, the
synthetic CMDs are populated with a number of stars much larger the
observed counterpart.

The distance modulus is assumed equal to 18.5 (see
e.g. \citealt{panagia91,pietr13}) and the total reddening,
E(B$-$V)$\approx 0.22$, is chosen by fitting the average color of the
UMS in the magnitude range $18-20$. Note that through this paper the
total reddening is defined as the composition of the Milky Way
foreground reddening, kept fixed at E(B$-$V)$\approx 0.07$ with
R$_{\mathrm{V}}=3.1$, and the local reddening. For the latter we used
the A$_{\lambda}$ values from \cite{demarchi16}, which are
specifically derived for the 30~Doradus environment.
An inspection of Figure \ref{4cmd} reveals that: 1) only ages in the
range 15-25 Myr allow one to match color and magnitude of the BL. The
synthetic BL of the 15 Myr isochrone fits the magnitude of the two
BSGs (although it is too blue), but it is clearly too bright on the
red side. On the other hand, the 20 Myr isochrone fits well the
magnitude of the three RSGs, but it fails to reproduce the ratio
RSGs/BSGs (the BL color extension is too short and its morphology
resembles a red clump); 2) Concerning the MSTO, even considering
photometric errors, none of the models is able to reproduce the
observed color spread in the magnitude range $16-18$. This is not
surprising given the presence of Be stars, which are thought to be
fast rotators surrounded by an out-flowing equatorial disk, likely
causing their infrared excess (see e.g. \citealt{pr03}); 3) only
isochrones between 25 Myr and 30 Myr allow one to reproduce the MSTO
luminosity (V$\sim\,16$; corresponding to a mass
$\approx \, 9\,$M$_{\odot}$ with PARSEC models), while the younger
ones are at least 1 mag brighter than the MSTO tip .

\begin{figure*}[t]
\centering \includegraphics[width=14cm]{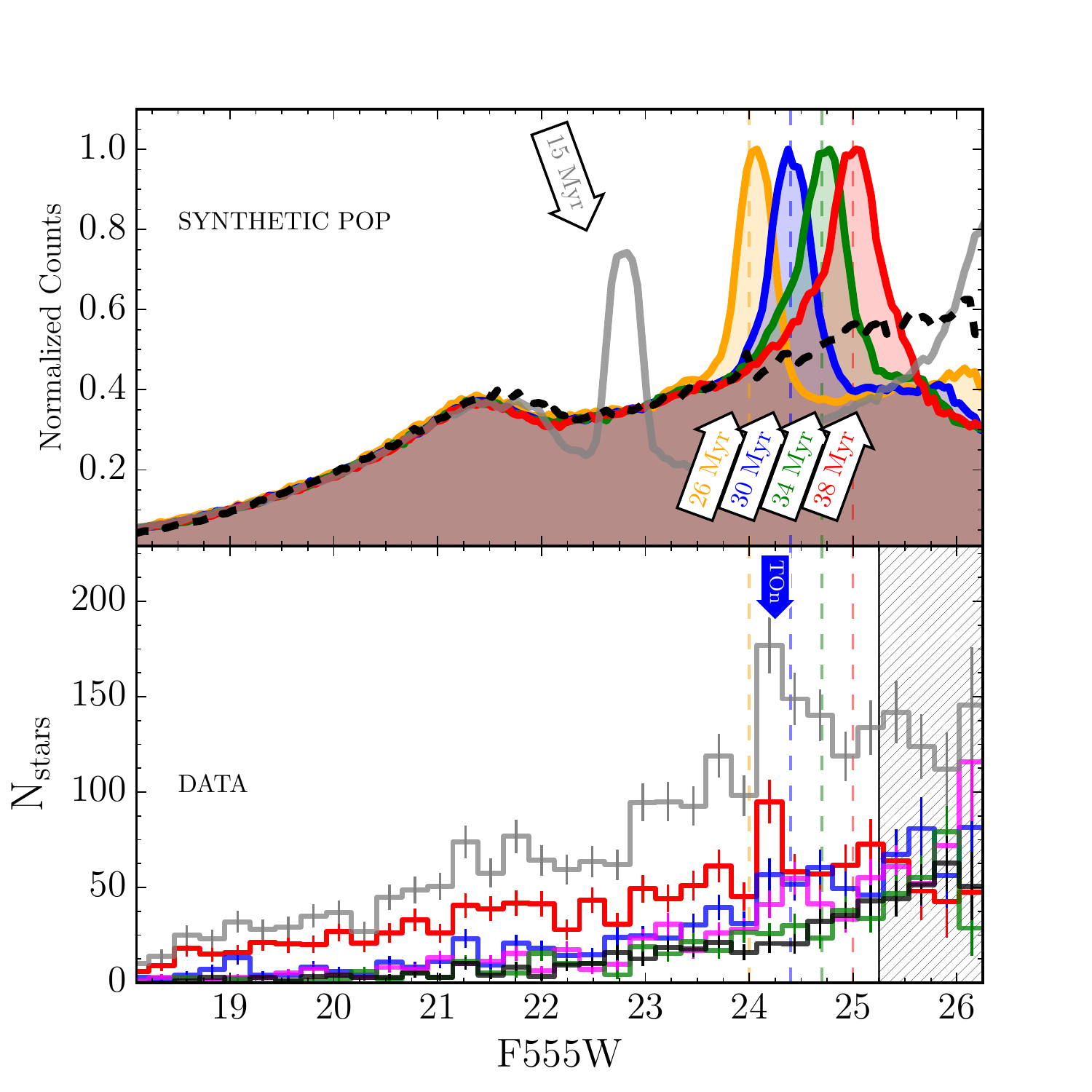}
\caption{Top panel: synthetic LFs for the labeled ages. The thick
  dashed line shows a synthetic zero age MS where the PMS phase has
  been artificially suppressed. Bottom panel: Observed LFs of stars in
  equal area annuli around H301's center (red for stars within 2.91
  pc, blue between 2.91 and 4.11 pc, magenta between 4.11 and 5.04 pc,
  green between 5.04 and 5.82 pc, black between 5.82 and 6.51 pc). The
  grey histogram is the sum of the three innermost LFs.  The magnitude
  of the TOn is also indicated.}
\label{LF_annuli} 
\end{figure*}

\emph{Low mass stars:} The PMS TOn is the point where the PMS joins
the MS. In H301 this phase is populated by low-mass stars below
$1.5\,$M$_{\odot}$. According to stellar evolution theory, at the
magnitude of the TOn, the luminosity function (LF) of a simple
population is characterized by a strong peak followed by a dip (see,
e.g., \citealt{cignoni10}). This behavior is clearly visible in the
top panel of Fig. \ref{LF_annuli} where synthetic populations of
different ages are shown (note that only photometric errors are
applied, while the incompleteness is not). The older the age of the
cluster, the fainter the LF TOn peak and the corresponding dip. Peak
and dip respectively reflect the steep dependence of stellar mass on
magnitude near the TOn and the following flattening below the TOn,
caused by the short evolutionary timescale of the PMS phase compared
to the MS. After the dip, the shape of the LF mimics the IMF, rising
with decreasing stellar mass. For comparison, the dashed line shows a
synthetic zero age MS where the PMS phase has been artificially
suppressed: as expected, without PMS there is no TOn peak/dip and LF
increases monotonically.

From a theoretical point of view, the dependence of TOn luminosity on
the PMS evolutionary times is a serious issue. However, while in the
first few Myr, PMS times are affected by several uncertainties like,
e.g., the initial conditions (in particular the initial radius), the
efficiency of convection, the initial abundance of deuterium and the
accretion rate (see, e.g., \citealt{baraffe02}, \citealt{tognelli11}),
at later times PMS tracks tend to converge. Indeed, the zero age MS
position for 1 M$_{\odot}$ does not show significant differences among
different authors (see Fig. 14 in \citealt{tognelli11}), with a
dispersion in $\log (\mathrm{L}/\mathrm{L}_{\odot})$ of $\approx 0.1$
dex. In terms of age, such an uncertainty corresponds to an age error
smaller than 3 Myr at 30 Myr.

Concerning the observational errors, two things limit the TOn
visibility in H301: incompleteness and field contamination. For ages
older than 15 Myr, at the distance of the LMC, a TOn is fainter than
23 in V magnitude, $\sim 24.5$ at 30 Myr. At these levels of
faintness, incompleteness can be severe, especially in the center of
H301. At odds with the massive stars, where the degree of
contamination is negligible, at faint magnitudes it can bias the age,
mimicking an older population.

The bottom panel of Figure \ref{LF_annuli} shows the observed LFs,
corrected for the incompleteness, of stars in five concentric annuli
of equal area around the center of H301 (in red, blue, magenta, green
and black stars between 0 and 2.91 pc, 2.91 and 4.11 pc, 4.11 and 5.04
pc, 5.04 and 5.82 pc, 5.82 and 6.51 pc, respectively). The hatched
area fainter than V=25.25 indicates where the incompleteness in the
innermost annulus drops below 50\%.

The TOn feature is visible as the narrow peak near V $\sim 24.25$ in
the LF of the innermost region (red histogram). At fainter magnitudes
the LF increases following the IMF, then it drops again at V$> 25.25$.
This is because below this limit the LF becomes increasingly affected
by photometric blends, whose net effect is to brighten the sample and
deplete the faint end. This happens for all annuli, but increases at
progressively fainter magnitudes as the distance from the cluster
center decreases. Indeed, the blue and magenta LFs (second and third
innermost annuli, respectively) show a general increase up to V=26,
reflecting the much more favorable incompleteness (below 50\% only at
V$>26$). However, the higher field contamination reduces the TOn
visibility, which is only noticeable as a broad peak in the magnitude
range $24-24.75$

Finally, the green and black LFs (the outermost annuli) show a smooth
increase, with no bumps in the range 24$-$25, as expected for the
average field of the LMC.

From a comparison by eye with the synthetic LFs we estimate the TOn
age to be between 26 and 30 Myr.

In the next Section we derive the most likely SFH compatible with the data.

\section{Quantitative derivation of H301's SFH}

To recover the most likely SFH we used the hybrid-genetic code SFERA
(Star Formation Evolution Recovery Algorithm), the statistical
approach described by \cite{cigno15}. Metallicity, binary fraction and
distance modulus are initially kept fixed at Z$=0.008$, 30\% and 18.5,
respectively, while age, reddening and field contamination are free
parameters. The SFH is parameterized in 40 contiguous steps of
duration 1 Myr between now and 40 Myr ago.  The best combination of
synthetic CMDs and field contamination (a template field is taken at
radii larger than 6 pc from the center) is searched by SFERA. In
SFERA, observational and model CMDs are binned in color and magnitude,
and the binning scheme is changed randomly.  The two 2D distributions
are then compared with a Poissonian $\chi^2$.

\begin{figure*}[t]
\centering \includegraphics[width=14cm]{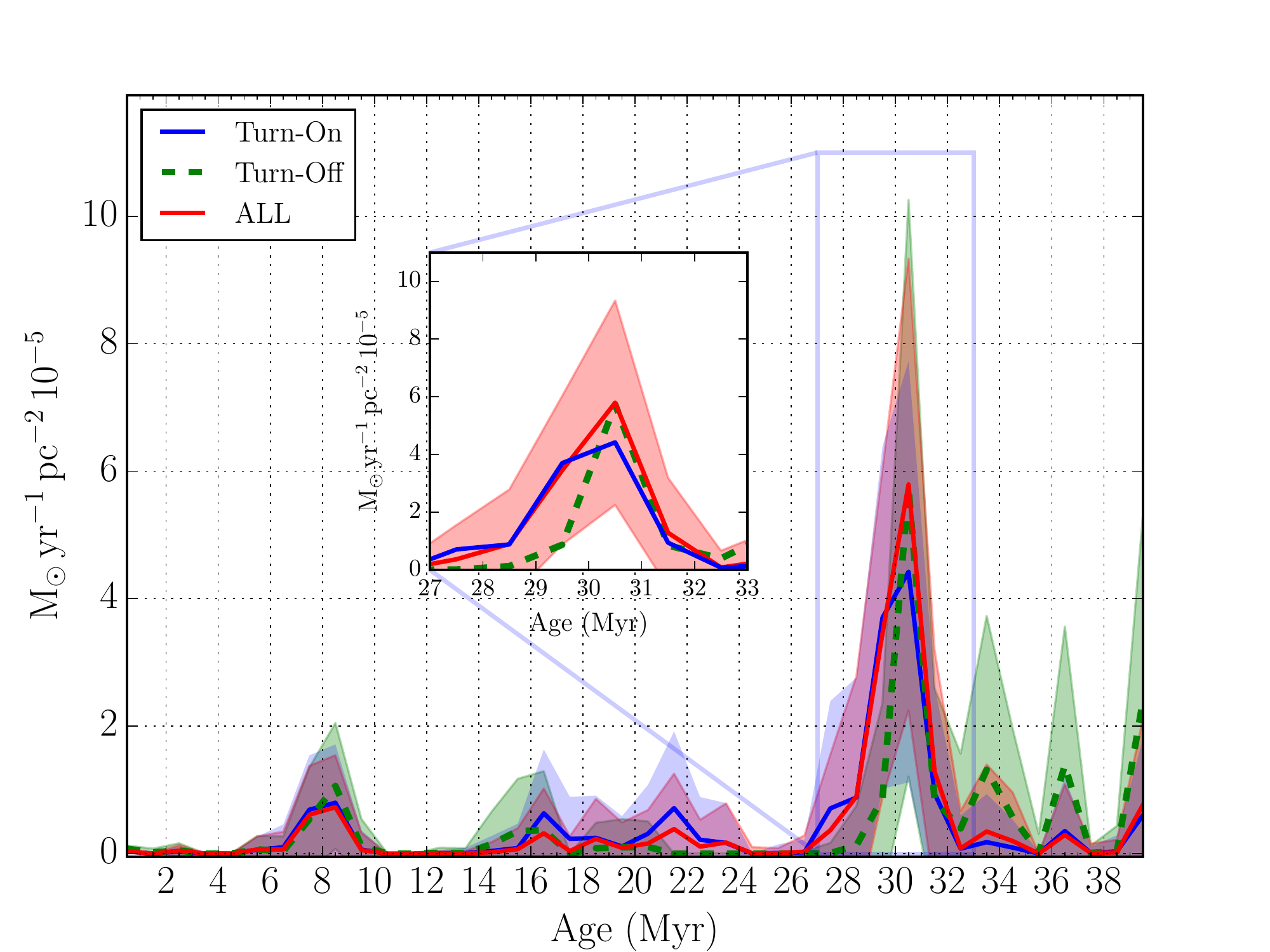}
\caption{SFHs of H301. Solid red, blue and dashed green lines
  corresponds to the best SFHs obtained using the entire CMD, TOn
  alone and MSTO/BL alone, respectively. Shaded regions represent
    the 1 $\sigma $ standard deviation from the best SFH.  The inset
  panel zooms around the main peak.}
\label{sfh} 
\end{figure*}

In order to reduce as much as possible the field contamination we only
used stars within 4 pc from H301's center. As a first step, we
recovered the SFH using massive stars (MSTO) and low-mass stars (TOn
stars) independently. The former includes all stars brighter than
F555W$=19$, the latter all stars fainter than F555W$=22$. Figure
\ref{sfh} shows the results, with the MSTO/BL solution (given the
paucity of BL stars, the age is largely driven by the MSTO) in dashed
green and PMS TOn one in solid blue. Finally, the red solid line shows
the SFH inferred using the entire CMD.
In all cases the most relevant peak is located in the range
$28.5$-$31.5$ Myr. The MSTO/BL solution shows a clear peak at
$30.5^{+1}_{-2}$ Myr, while the TOn peak is slightly broader on the
young side, suggesting that the MSTO age is better defined (not
surprising given the much smaller photometric error affecting this
phase). The net result is that, within the errors, MSTO and TOn ages
are in excellent agreement. Combining the two features leads to the
red solid line solution, which resembles more the MSTO solution, but
with smaller uncertainties. Hereafter we refer to this solution as
the best SFH for H301.

At first glance our conclusions seem at odds with the results derived
by \cite{naylor09}, who studied the age for a selection of clusters
and associations younger than 100 Myr. They found that the ages based
on PMS isochrones are 1.5-2.0 times shorter than the ``nuclear'' MSTO
ages. However, our PMS age is mostly based on the TOn luminosity, and
not on the PMS stars, whose color is affected by several uncertainties
(see e.g. \citealt{henne}). Another possible source of error is the
use of different sets of models to study different phases: we do not
have this kind of problem because in this work we used the PARSEC
models which follow the entire evolution from the PMS to the post-MS.

\begin{figure*}[t]
\centering \includegraphics[width=12cm]{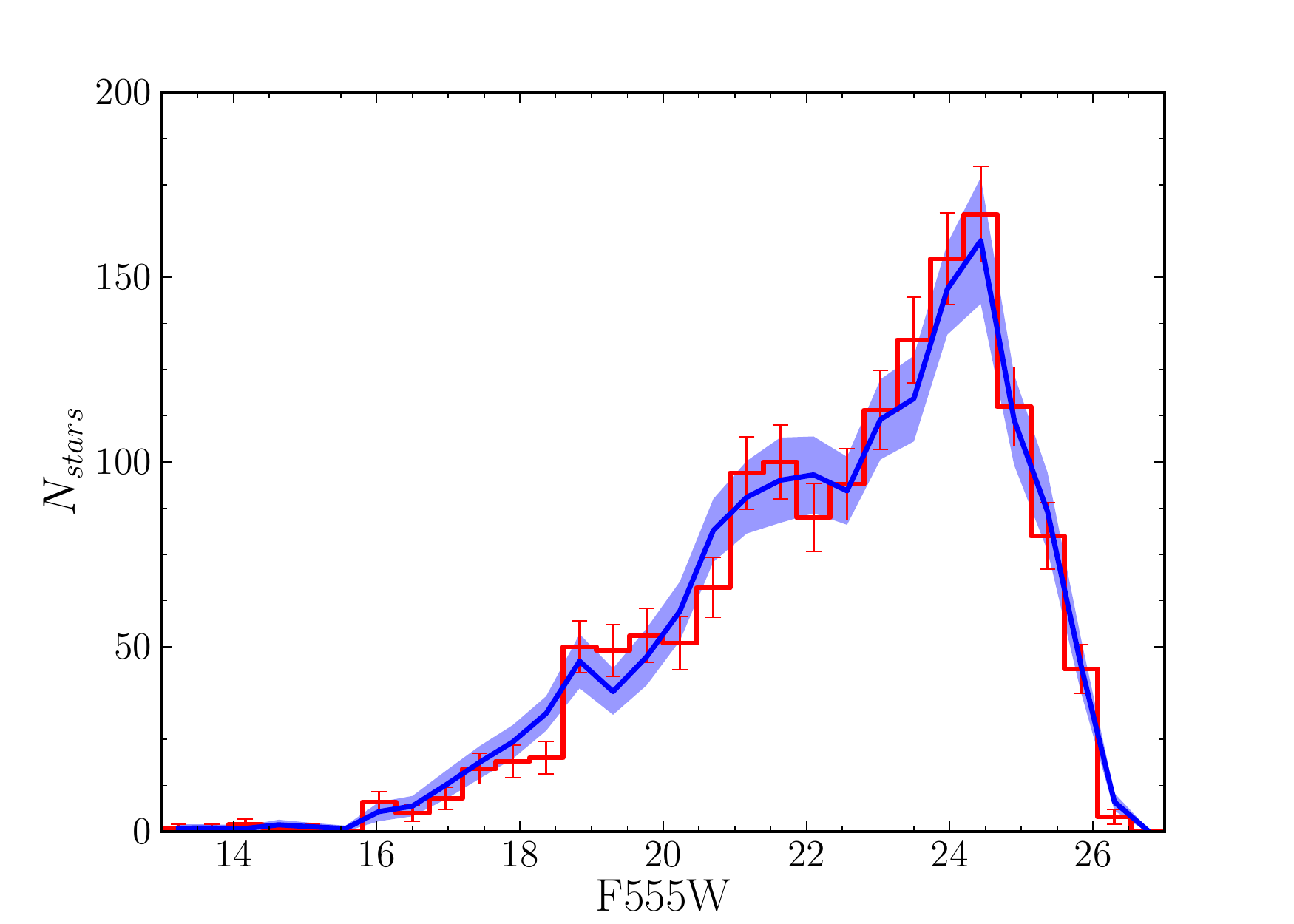}
\caption{Synthetic LF generated with the most likely SFH compared with
  the observational LF. }
\label{lf2} 
\end{figure*}

Fig. \ref{lf2} shows the synthetic LF (blue line) corresponding to the
best SFH compared with the observed counterpart (red line). Error bars
in the data are the square root of the counts, while the $1\,\sigma$
uncertainty in the model is indicated with a blue band. The quality of
the fit is excellent and most of the differences are within the
errors.
We found a best-fitting E(B$-$V) of 0.22-0.24 mag with a $1\,\sigma$
dispersion of 0.04 mag. In this case the dispersion is not the actual
error, but the spread (differential reddening) needed to match the MS
width.

It is worth to notice that the inferred absolute ages are expected to
change according to the adopted distance modulus. Indeed a shorter
distance would increase the cluster age. However, in the range 20-30
Myr, both the MSTO and TOn magnitudes change with age by
$\approx 0.1\, \mathrm{mag}/\mathrm{Myr}$, hence the relative age
MSTO/TOn remains unchanged.

In order to derive the initial total mass and number of supernovae
Type II exploded in H301 we ran 1000 Monte Carlo simulations
normalized to the observed number of stars (decontaminated with an
external field) brighter than F555W $=20$. Adopting the peak age
$30.5$ Myr, we calculated that the initial total mass of H301 was
$\approx 8800\,\pm 800$ M$_{\odot}$\footnote{Corrected for stars
  residing outside the 4 pc radius (about 15\%).} and that $52\pm 9$
supernovae Type II\footnote{Assuming that all stars above about 20
  M$_\odot$ produce black holes.} exploded. This result is also
supported by observations (see GC00 and references therein;
\citealt{lopez11}), that place H301 within an high-velocity expanding
shell, caracterized by multiple centers and filled with diffuse X-ray
emission. Moreover, \cite{walborn97} classified an object, WB9 (Be1 in
GC00; indicated with a green circle in Fig. \ref{chart}), as a
spectroscopic binary with a compact companion, possibly the stellar
remnant of a supernova event.

Before closing this section, it is important to discuss the impact of
the assumed binary recipe and metallicity on the recovered SFH.

\subsection{Binaries}

Although our findings
do not critically depend on the adopted binary fraction and mass
ratio, a very high binary fraction (above 50\%; \citealt{sana13}) and
flatter mass ratio q (our binary prescription favors low q binaries),
as found for O-stars, can affect the recovered SFH. In practice, the
effect of equal mass binaries is to make stars appear 0.75 mag
brighter, and not taking it into account leads to an underestimate of
the age. Since our SFHs are derived disfavouring q=1 massive binaries
we may indeed interpret the observed brightnesses underestimating the
age if the binaries have mostly the same mass.

However, B-type stars may behave differently. \cite{dunstall15}
studied the multiplicity of 408 B-type stars observed in different
regions (NGC~2070, NGC~2060, Hodge~301, SL~639) of 30~Doradus with
multi-epoch spectroscopy from the VLT-FLAMES Tarantula Survey
(VFTS). Although they found an average binary fraction of about 58\%,
close to the O-stars multiplicity by \cite{sana13}, the intrinsic
binary fraction in H301 was found to be remarkably low and around 20\%
(8\% detected, with a detection probability of $40 \pm
10$\%).
Concerning the mass ratio, they found a distribution of q favoring
low-mass companions (f(q)$\propto q^{-2.8}$).

\begin{figure*}[t]
\centering \includegraphics[width=14cm]{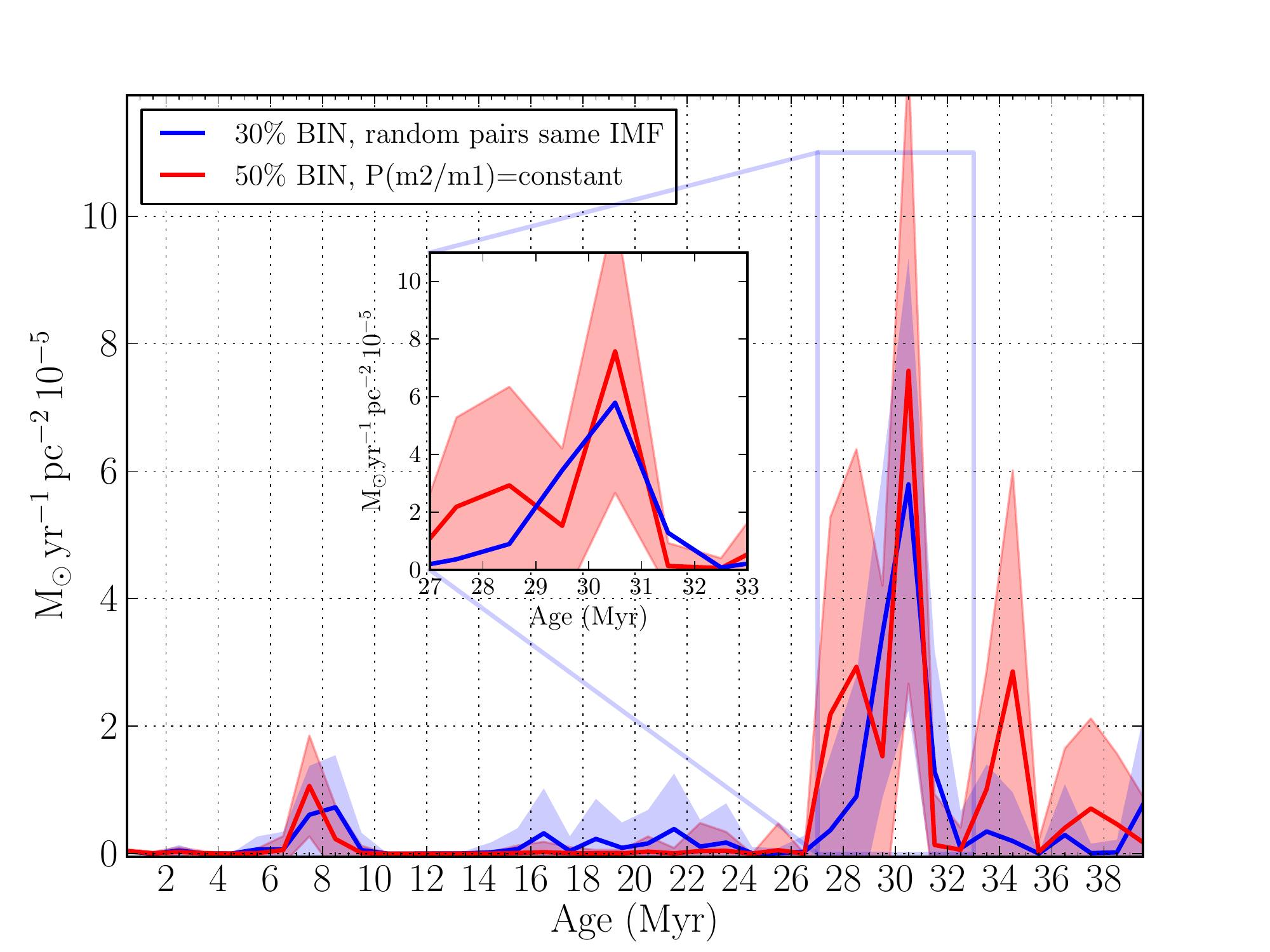}
\caption{Red slope: SFH recovered adopting 50\% of binaries and
  constant mass ratio. }
\label{bin} 
\end{figure*}

Although our binary prescription is consistent with
\possessivecite{dunstall15} finding, we also tested the hypothesis of
a binary population similar to that found by \cite{sana13} for O-type
stars. Figure \ref{bin} shows the SFH recovered assuming a binary
fraction of 50\% and a uniform distribution of q. Overall, the result
is qualitatively similar to the standard SFH, except the peak that is
slightly broader.

Finally, in the literature there is no evidence of a universal mass
ratio holding from massive to low-mass stars (seey 
e.g. \citealt{ward15}). However, the similarity of our inferred MSTO
and TOn ages suggests that binary populations of high and low-mass
stars are not dramatically different.

\subsection{Metallicity}

\begin{figure*}[t]
\centering \includegraphics[width=14cm]{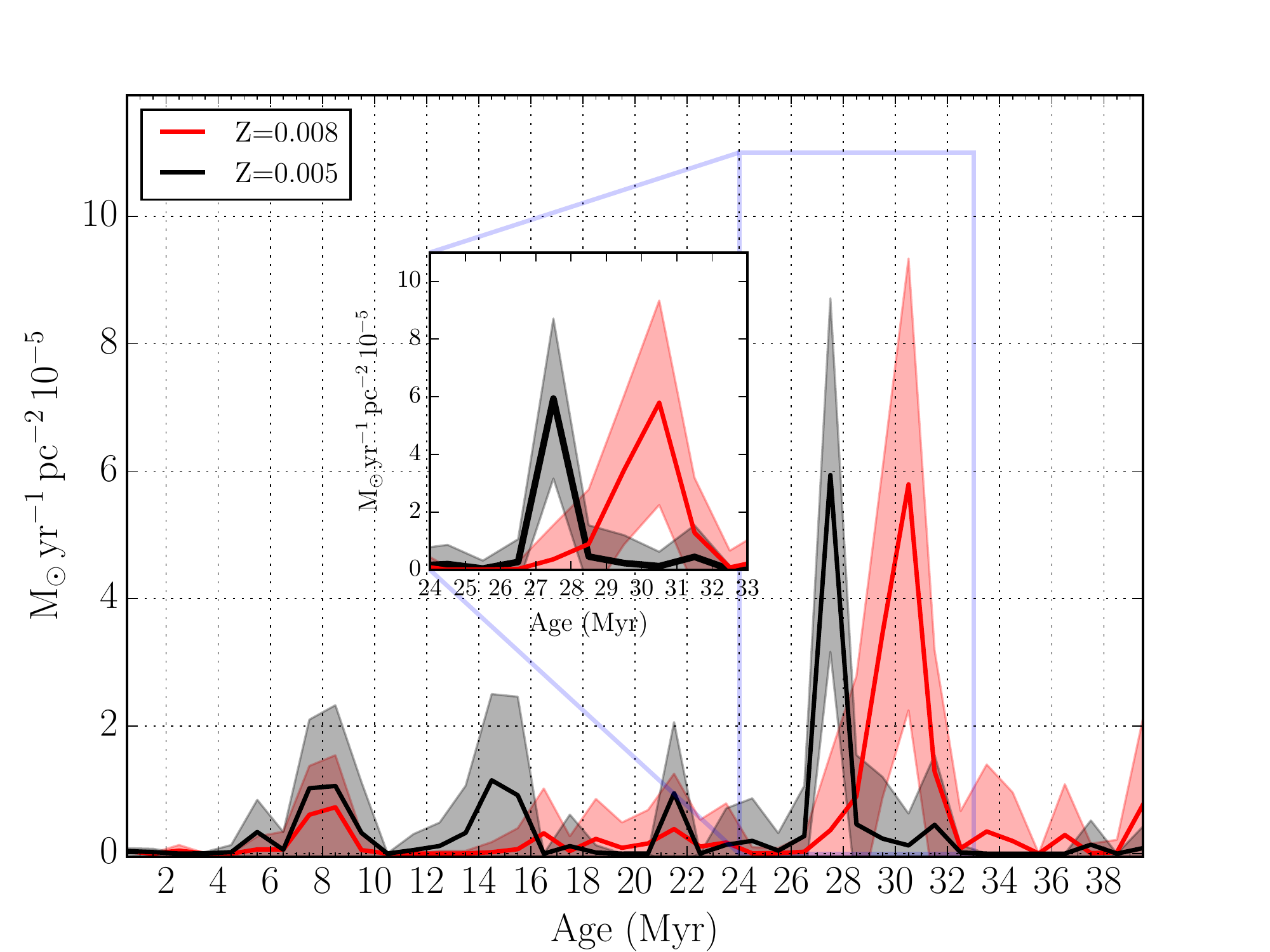}
\caption{Recovered SFH for the metallicity Z=0.005 (black line) compared
  to the Z=0.008 one (red line). }
\label{sfr_feh} 
\end{figure*}

The empirical evidence of a metallicity spread in the young
populations of the LMC (see e.g. \citealt{luck98},
\citealt{romaniello08}) makes Hodge's metallicity inherently
uncertain. In order to test the impact of a lower metallicity on the
final age, we re-recovered the SFH using PARSEC models with Z=0.005
(about $1\,\sigma$ away from the mean value of Luck's sample). The
result is shown in Fig. \ref{sfr_feh}. The lower metallicity reduces
the peak age by about 3 Myr. In this case, the best age is $27.5\pm 1$
Myr.

This difference can be explained by the dependence of MSTO and TOn
luminosities on metallicity and helium abundance (PARSEC isochrones at
Z=0.005 have slightly lower helium abundance than isochrones at
Z=0.008).  First, the lower the metallicity, the shorter the
evolutionary timescale of a star of a given mass during the major
evolutionary phases. For this reason, at a fixed age, MSTO and TOn
masses are lower at Z=0.005 than at Z=0.008, and stars of lower masses
are also fainter.  However, as secondary effect, a decrease of
metallicity shifts the isochrone to higher luminosities, whereas a
decrease of helium has the opposite effect. The combined effect of
these changes decreases MSTO and TOn luminosities of 0.1-0.2 mag,
mimicking an older isochrone. The resulting mean reddening is also
higher by about 0.02 mag (E(B-V)=0.24 mag), which causes a further
luminosity decrease of about 0.1 mag.

The new peak age, 27.5 Myr, leaves the initial total mass of H301
almost unchanged, whereas the number of supernovae Type II, $46\pm 8$,
is slightly lower than in the Z=0.008 case.

\begin{figure}[t]
\centering \includegraphics[width=10cm]{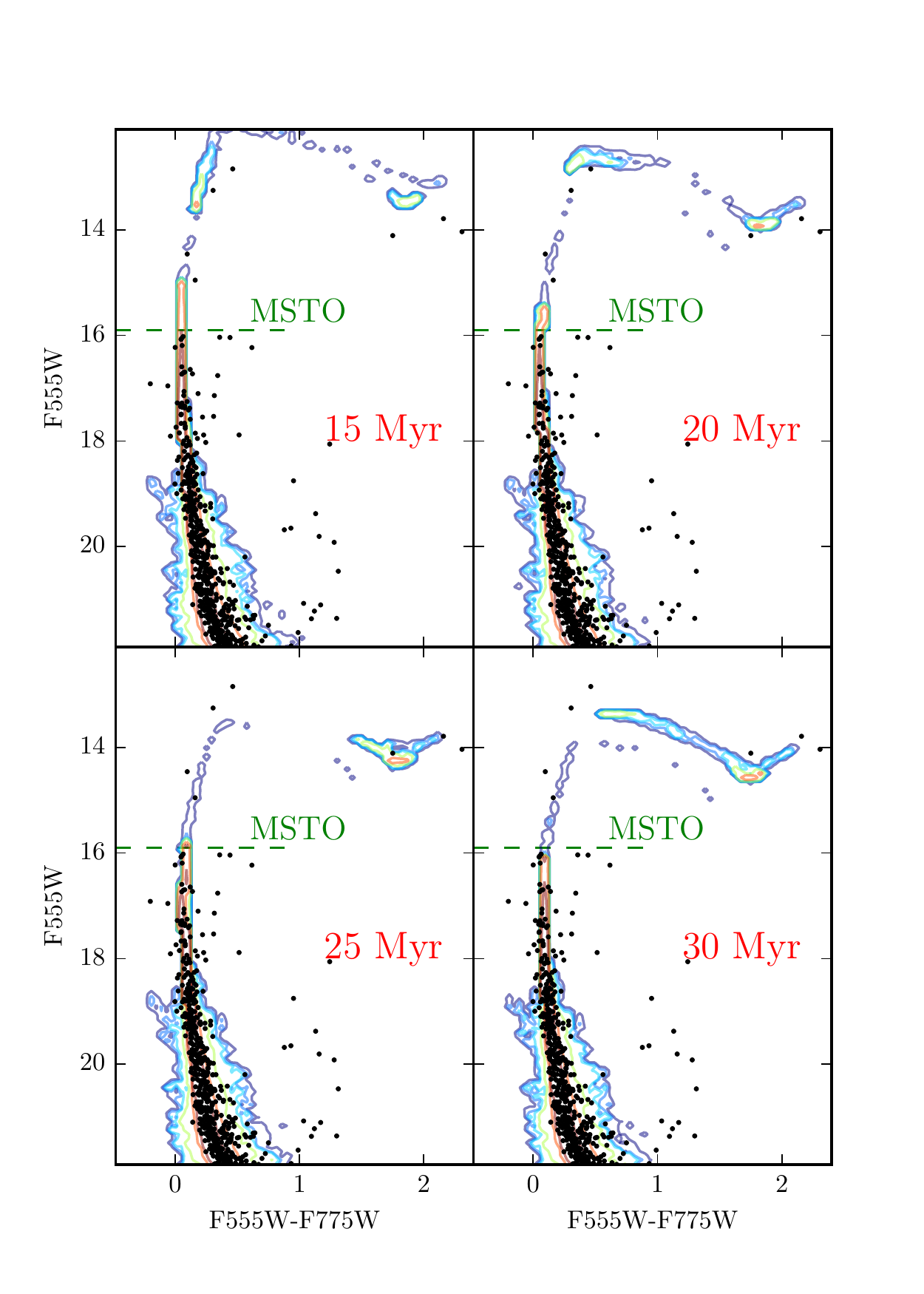}
\caption{The same as in Fig. \ref{4cmd}, but the synthetic CMDs are computed for
  Z=0.005 instead of Z=0.008.}
\label{4cmd_fe05} 
\end{figure}

Another effect of the lower metallicity is to increase the luminosity
of the BL phase. This is clearly visible in Fig. \ref{4cmd_fe05}. In
contrast to Z=0.008 isochrones (see Fig. \ref{4cmd}), the 20 Myr BL is
long enough to reach the color of the two BSGs, but the red side is
now brighter than the three RSGs. On the other hand, the 25 Myr
isochrone plays the same role of the 20 Myr isochrone at Z=0.008,
fitting well the RSGs luminosity but failing to reproduce the observed
ratio RSGs/BSGs. The entire 30 Myr BL is too faint. In conclusion,
while no one model is able to reproduce extension and luminosity of
the BL at the same time, the 25 Myr is the one that better represents
the observed BL, getting closer to the MSTO/TOn best ages, which are
now only 2.5 Myr apart.

In summary, given the uncertainty in the metallicity of H301, we
conclude that the best fitting age is between 26.5 and 31.5 Myr, while
the predicted number of supernovae Type II is between 38 and 61.

\section{Comparison with the literature}

Compared to the literature, our lower limit on the age of H301,
$\sim 26.5$ Myr, is slightly older than the photometric estimate of
GC00, who found $20\pm 5$ and $25\pm 5$ Myr using Geneva
(\citealt{schaerer93}) and Padova models (\citealt{bertelli94})
respectively. GC00 required their isochrone fit to match the position
of the blue and red supergiants and omitted stars with H$\alpha$
excess. This difference is probably caused by the different versions
of the isochrones and different stellar phases used in our
analysis. Indeed, if we limit our analysis to the BL phase only (see
Fig. \ref{4cmd} and \ref{4cmd_fe05}), our best estimate of the age
would drop down to 20-25 Myr, in good agreement with GC00. GC00's
estimate of the total reddening (E(B-V)$=0.28\pm 0.05$ mag) is
compatible with our estimate.

Concerning the mass of H301, GC00 found a present day mass of
4882$\pm$247 M$_\odot$ between 0.4 and 12 M$_\odot$, corresponding to
an initial mass of $\approx 6000 \pm 525$ M$_{\odot}$ when
extrapolated with a Salpeter IMF above 12 and up to 120
M$_{\odot}$. Most of the difference from our estimate stems from the
different age and IMF. If we were to adopt a Salpeter IMF in our
  models down to 0.4 M$_{\odot}$, our mass estimate would drop to
  $\sim 7200 \pm 700$ M$_{\odot}$. In addition, if we were to adopt a
  20 Myr isochrone (as adopted by GC00) instead of a 30.5 Myr ago, our
  result would drop to $6600\pm 400$ M$_{\odot}$, in excellent
  agreement with GC00's estimate. Our estimated number of supernovae,
  between 38 and 61, is in good agreement with GC00's rate
  ($41\pm 7$).

  Our best age is higher than that found by Evans (15$\pm$5 Myr), who
  used high resolution spectroscopy to reconstruct temperature and
  luminosity for a sample of B-stars within 4.9 pc from H301's
  center. Part of the discrepancy can be ascribed to the adopted
  stellar models from \cite{brott11}, whose timing is faster than our
  Padova models of the same metallicity: we found that a Padova
  isochrone of 30 Myr has the same MSTO luminosity of Brott et
  al. isochrone of 25 Myr. The rest of the discrepancy might then be
  attributed to two other potential effects; the choice MSTO magnitude
  in the present paper (Fig. \ref{cmd} and see the accompanying
  discussion in section 2), and the transformation from a CMD to the
  HRD in Evans et al.  It is clear from Fig. 5 of Evans et al. that
  they include the fainter of the two PMSG stars in their main
  sequence group, and from our Fig.6 one might argue that raising the
  MSTO to V$=\,15$ would give an age of ~20 Myr, in better agreement
  with Evans et al.  However a more serious issue perhaps is that
  Evans et al. derive luminosities from the stars' effective
  temperatures that imply there are several stars in H301 with masses
  between 12 and 15 solar masses, significantly in excess of the 9-10
  M$_{\odot}$ implied by an age of 26.5-31.5 Myr of the PARSEC
  models. The brightest of these B-stars, particularly those in the
  PMSG, might be explained as blue stragglers, being the products of
  binary evolution. The present low fraction of binaries in H301,
  discussed in the previous section, may simply reflect that many of
  these have already interacted and formed binary products as
  discussed by \cite{schneider14} producing blue straggler stars that
  are among the brightest blue stars in H301.

Finally, it is important to remark that a 15 Myr age would be strongly
ruled out by the PMS TOn. As shown in Fig. \ref{LF_annuli}, a TOn of
15 Myr (grey slope) would be at least 1 mag brighter than the present
TOn, and hence less affected by photometric errors and
incompleteness. If any significant excess of counts at
V$\approx 22.5-23$ were present, it would be clearly
detected. Moreover, the 15 Myr age is disfavored by the MSTO too: as
shown in Fig. \ref{4cmd}, the isodensity contours of the 15 Myr model
show a continuity of counts up to the two PMSG stars, while no stars
are observed in the magnitude range 15-16.

We further note, that a search for PMS stars based on the H$\alpha$
excess emission (e.g. \citealt{demarchi10,demarchi11}) over the whole
HTTP area has revealed an overdensity of about 120 PMS stars within 4
pc of H301, with an average age of $\sim 28$\, Myr (De Marchi,
Panagia, Sabbi, et al., in preparation), hence in excellent agreement
with our estimate.

Although of low significance, a mild activity in the range 6-8 Myr is
also predicted by most of our solutions. Indeed, for V$> 23$, a few
stars up to 0.5 mag redder than the MS are visible at all radii (see
Fig. \ref{field}), hence suggesting that if a more recent episode of
star formation took place this did not stem necessarily from H301 but
it involved a larger region, as generated in a more diffuse
environment or from a dissolved cluster. An inspection of the
H$\alpha$ flux reveals that some of these red objects have H$\alpha$
excess, but their signal-to-noise ratio is too low ( $1-2\, \sigma$)
for a firm conclusion.  Another possibility is that a few stars in the
region suffer from much higher reddening, mimicking a PMS
population.

\section{Conclusions}

From comparison of the observed CMD with simulations based on stellar
evolutionary models we derive in a self-consistent way the age
distribution and reddening of Hodge~301, a young cluster located in
the 30~Doradus region. Thanks to the photometric capabilities of the
HTTP data-set we have detected the PMS TOn for the first time. The
peak age we derive from fitting this feature and the MSTO, between
26.5 and 31.5 Myr ($30.5^{+1}_{-2}$ Myr using Z=0.008, $27.5\pm 1$ Myr
using Z=0.005), confirms that Hodge~301 is much older than the bulk of
the stars in NGC~2070, the most active region of 30~Doradus, but only
slightly older than its oldest stars ($\approx 20$ Myr;
\citealt{cigno15}). For a fixed metallicity, the resulting age spread,
$\approx 1-3$ Myr, is of the order of the age uncertainty as expected
from photometric errors only, hence it is difficult to conclude
whether the spread reflects a real prolonged SF.

The inferred PMS TOn age is consistent with the age derived from the
MSTO and a few Myr older than the age derived from fitting the
luminosity of the post-MS stars. In particular, while none of the
models can reproduce BL extension and luminosity at the same time,
fitting the three RSGs leads to an age between 20 and 25 Myr. However,
post-MS theoretical models for intermediate/massive stars are very
uncertain. As shown in, e.g., \citealt{tan14} mass loss and
core-convective overshooting can greatly affect the BL color
extension. Indeed, models tend to predict a clear separation in the
CMD between MS and BL stars, while no such gap is seen in several
extragalactic young massive star clusters (see
e.g. \citealt{larsen11}) and dwarf galaxies (see
e.g. \citealt{tang14,tang16} ). Given this uncertainty, we are
inclined to favor a cluster age that is mainly derived from fitting
the MSTO and PMS TOn. More in general, while the absolute age of H301
does depend on the stellar models adopted, the age difference between
H301 and NGC2070 (\citealt{cigno15}) is robust, since the analysis is
done with the same technique, stellar models and clock (PMS TOn).

Finally, it is intriguing that the MSTO/PMS TOn age estimate is also
older than the spectroscopic age derived by \cite{evans15}. Part of
the discrepancy could be attributed to the presence of blue
stragglers. However, unless H301 hosts an unusual number of these
objects, the discrepancy could indicate problems in current stellar
evolutionary models of massive stars.

Other interesting results are: 1) H301's total stellar mass is
$\approx 8800\,\pm 800$ M$_{\odot}$; 2) the total reddening E(B$-$V)
is $\approx 0.22-0.24$ mag, with a dispersion of 0.04 mag; 3) between
  38 and 61 supernovae Type-II exploded in the region.

From the point of view of the Tarantula Nebula, the old age of H301,
several Myr older than the nearby and massive NGC~2070, and its high
supernovae activity, along with the fact that not older clusters are
visible in the region, could suggest that the onset of H301 sparked
the formation of NGC~2070.

\acknowledgments 

We would like to thank Mario Gennaro, Chris Evans and Pier Giorgio
Prada Moroni for helpful comments and discussions. D.A.G. kindly
acknowledges financial support by the German Research Foundation (DFG)
through grant GO\,1659/3-2. EKG gratefully acknowledges funding from
Sonderforschungsbereich ``The Milky Way System'' (SFB 881) of the
German Research Foundation (DFG), especially via subproject B5.


\begin{thebibliography}{}

\bibitem[Baraffe et al.(2002)]{baraffe02} Baraffe, I., Chabrier, G., Allard, F., \& Hauschildt, P.~H.\ 2002, \aap, 382, 563 


\bibitem[Bertelli et al.(1994)]{bertelli94} Bertelli, G., Bressan, A.,
  Chiosi, C., Fagotto, F., \& Nasi, E.\ 1994, \aaps, 106,

\bibitem[Bressan et al.(2012)]{bressan12} Bressan, A., Marigo, 
P., Girardi, L., et al.\ 2012, \mnras, 427, 127 

\bibitem[Brott et 
al.(2011)]{brott11} Brott, I., de Mink, S.~E., Cantiello, M., et al.\ 2011, \aap, 530, A115 

\bibitem[Caffau et al.(2011)]{caffau11} Caffau, E., Ludwig, H.-G., Steffen, M., Freytag, B., \& Bonifacio, P.\ 2011, \solphys, 268, 255 



\bibitem[Cignoni et al.(2010)]{cignoni10} Cignoni, M., Tosi, M., Sabbi, E., et al.\ 2010, \apjl, 712, L63 


\bibitem[Cignoni et al.(2015)]{cigno15} Cignoni, M., Sabbi, E., 
van der Marel, R.~P., et al.\ 2015, \apj, 811, 76 

\bibitem[De Marchi et al.(2010)]{demarchi10} De Marchi, G., Panagia, N., \& Romaniello, M.\ 2010, \apj, 715, 1 

\bibitem[De Marchi et al.(2011)]{demarchi11} De Marchi, G., Paresce, F., Panagia, N., et al.\ 2011, \apj, 739, 27 


\bibitem[De Marchi et al.(2016)]{demarchi16} De Marchi, G., Panagia, N., Sabbi, E., et al.\ 2016, \mnras, 455, 4373 

\bibitem[Dunstall et al.(2015)]{dunstall15} Dunstall, P.~R., Dufton,
  P.~L., Sana, H., et al.\ 2015, \aap, 580, A93




\bibitem[Evans et al.(2015)]{evans15} Evans, C.~J., Kennedy, M.~B.,
  Dufton, P.~L., et al.\ 2015, \aap, 574, A13

\bibitem[Grebel et al.(1996)]{grebel1996} Grebel, E.~K., Roberts, W.~J., \& Brandner, W.\ 1996, \aap, 311, 470 







\bibitem[Grebel(1997)]{grebel1997} Grebel, E.~K.\ 1997, \aap, 317, 448


\bibitem[Grebel \& Chu(2000)]{grebel00}
Grebel, E.K., Chu, Y.-H., 2000, AJ, 111, 787




\bibitem[H{\'e}nault-Brunet et al.(2012)]{henault12}
  H{\'e}nault-Brunet, V., Evans, C.~J., Sana, H., et al.\ 2012, \aap,
  546, AA73

\bibitem[Hennekemper et al.(2008)]{henne} Hennekemper, E., Gouliermis, D.~A., Henning, T., Brandner, W., \& Dolphin, A.~E.\ 2008, \apj, 672, 914-929 


\bibitem[Hodge(1988)]{ho88} Hodge, P.\ 1988, \pasp, 100, 
1051 

\bibitem[Hunter et al.(2008)]{hunter08} Hunter, I., Lennon, D.~J., Dufton, P.~L., et al.\ 2008, \aap, 479, 541 


\bibitem[Kroupa(2001)]{kroupa01} Kroupa, P.\ 2001, \mnras, 322, 
231 

\bibitem[Larsen et al.(2011)]{larsen11} Larsen, S.~S., de Mink, S.~E., Eldridge, J.~J., et al.\ 2011, \aap, 532, A147 




\bibitem[Lopez et al.(2011)]{lopez11} Lopez, L.~A., Krumholz, M.~R., Bolatto, A.~D., Prochaska, J.~X., \& Ramirez-Ruiz, E.\ 2011, \apj, 731, 91 



\bibitem[Lortet \& Testor(1991)]{lortet91} Lortet, M.-C., \& Testor, G.\ 1991, \aaps, 89, 185 

\bibitem[Luck et al.(1998)]{luck98} Luck, R.~E., Moffett, 
T.~J., Barnes, T.~G., III, \& Gieren, W.~P.\ 1998, \aj, 115, 605 


\bibitem[McEvoy et al.(2015)]{mcevoy15} McEvoy, C.~M., Dufton, P.~L., Evans, C.~J., et al.\ 2015, \aap, 575, A70 

\bibitem[McGregor \& Hyland(1981)]{mcgregor81} McGregor, P.~J., \& Hyland, A.~R.\ 1981, \apj, 250, 116 

\bibitem[Melnick(1985)]{melnick85} Melnick, J.\ 1985, \aap, 153, 235 


\bibitem[Mendoza V.~\& G{\'o}mez(1973)]{mendoza73} Mendoza V., E.~E., \& G{\'o}mez, T.\ 1973, \pasp, 85, 439 


\bibitem[Naylor(2009)]{naylor09} Naylor, T.\ 2009, \mnras, 399, 432 



\bibitem[Panagia et al.(1991)]{panagia91} Panagia, N., Gilmozzi, 
R., Macchetto, F., Adorf, H.-M., \& Kirshner, R.~P.\ 1991, \apjl, 380, L23 

\bibitem[Pietrzy{\'n}ski et al.(2013)]{pietr13} 
Pietrzy{\'n}ski, G., Graczyk, D., Gieren, W., et al.\ 2013, \nat, 495, 76 


\bibitem[Porter \& Rivinius(2003)]{pr03} Porter, J.~M., \& Rivinius,
  T.\ 2003, \pasp, 115, 1153 

\bibitem[Romaniello et al.(2008)]{romaniello08} Romaniello, M., Primas, F., Mottini, M., et al.\ 2008, \aap, 488, 731 




\bibitem[Sabbi et al.(2012)]{sabbi12} Sabbi, E., Lennon, D.~J., Gieles, M., et al.\ 2012, \apjl, 754, LL37 


\bibitem[Sabbi et al.(2015)]{sabbi15}  Sabbi, E., Lennon, D.~J., Anderson, J., et al.\ 2016, \apjs, 222, 11 



\bibitem[Sana et al.(2013)]{sana13} Sana, H., de Koter, A., de Mink, S.~E., et al.\ 2013, \aap, 550, A107 



\bibitem[Schaerer et 
al.(1993)]{schaerer93} Schaerer, D., Meynet, G., Maeder, A., \& Schaller, G.\ 1993, \aaps, 98, 523 

\bibitem[Schneider et al.(2014)]{schneider14} Schneider, F.~R.~N., Izzard, R.~G., de Mink, S.~E., et al.\ 2014, \apj, 780, 117 



\bibitem[Tang et al.(2014)]{tan14} Tang, J., Bressan, A., Rosenfield, P., et al.\ 2014, \mnras, 445, 4287 



\bibitem[Tang et al.(2014)]{tang14} Tang, J., Bressan, A., Rosenfield,
  P., et al.\ 2014, \mnras, 445, 4287


\bibitem[Tang et al.(2016)]{tang16} Tang, J., Bressan, A., Slemer, A., et al.\ 2016, \mnras, 455, 3393 


\bibitem[Tognelli et al.(2011)]{tognelli11} Tognelli, E., Prada Moroni, P.~G., \& Degl'Innocenti, S.\ 2011, \aap, 533, A109 



\bibitem[Vink et al.(2010)]{vink10} Vink, J.~S., Brott, I., Gr{\"a}fener, G., et al.\ 2010, \aap, 512, L7 

\bibitem[Walborn et al.(1989)]{walborn89} Walborn, N.~R., Prevot, M.~L., Prevot, L., et al.\ 1989, \aap, 219, 229 


\bibitem[Walborn 
\& Blades(1997)]{walborn97} Walborn, N.~R., \& Blades, J.~C.\ 1997, \apjs, 112, 457 


\bibitem[Ward-Duong et al.(2015)]{ward15} Ward-Duong, K., 
Patience, J., De Rosa, R.~J., et al.\ 2015, \mnras, 449, 2618




\end{thebibliography}
\end{document}